\documentclass[preprint,3p]{elsarticle}
\usepackage[toc,page,title,titletoc,header]{appendix}
\usepackage{stmaryrd}
\SetSymbolFont{stmry}{bold}{U}{stmry}{m}{n}
\usepackage{amsmath}
\usepackage{amssymb}
\usepackage{amsthm}
\usepackage{tabularx}
\usepackage{subfigure}
\usepackage{epsfig}
\usepackage{indentfirst}
\usepackage{epstopdf}
\usepackage{cases}
\biboptions{sort&compress}
\usepackage{graphicx}
\usepackage{color}
\usepackage{float}
\usepackage{multirow}

\newcommand{\abs}[1]{\lvert#1\rvert}

\renewcommand{\vec}[1]{\mathbf{#1}}

\def\subFV{\scriptscriptstyle{FV}}
\def\subFS{\scriptscriptstyle{FS}}
\def\subVS{\scriptscriptstyle{VS}}

\newtheorem{prop}{Proposition}[section]

\journal{J. Comput. Applied. Math}
\numberwithin{equation}{section}

\begin{document}

\begin{frontmatter}

\title{A sharp-interface model and its numerical approximation for solid-state dewetting with axisymmetric geometry}

\author{Quan Zhao}
\address{Department of Mathematics, National University of
Singapore, Singapore, 119076}
\ead{quanzhao90@u.nus.edu}


\begin{abstract}
Based on the thermodynamic variation, we rigorously derive the sharp-interface model for solid-state dewetting on a flat substrate in the form of cylindrical symmetry. The governing equations for the model belong to fourth-order geometric curve evolution partial differential equations, with proper boundary conditions such that the total volume of the system is conserved and the total energy is dissipative during the time evolution. We propose a variational formulation for the sharp-interface model and then apply the parametric finite element method for solving it efficiently. Extensive numerical simulation results are presented lastly to demonstrate the morphological characteristics for solid-state dewetting. 
\end{abstract}



\begin{keyword}
Solid-state dewetting, axisymmetric geometry,  thermodynamic variation, surface diffusion flow, contact line migration, parametric finite element method (PFEM).
\end{keyword}

\end{frontmatter}

\section{Introduction}

A drop of water on a leaf or a soap bubble in the air tends to form near-spherical shape, as the sphere has the smallest surface area to volume ratio. This phenomenon, known as surface tension, results from the cohesive forces among the liquid molecules and drives the system to move towards the state with lower surface energy. In micro and nano-scale, surface tension effects can also be observed in solids. The solid thin films supported by rigid substrates are typically unstable and demonstrate high instability even if they are at a temperature below their melting points. They could dewet or agglomerate to form small particles on the substrate, and exhibit complex morphological evolutions even below their melting temperature~\cite{Jiran90, Jiran92,Ye10a,Ye10b,Ye11a,Ye11b,Rabkin14}. During this process, the thin film remains in the solid state, thus it is called solid-state dewetting~\cite{Thompson12, Leroy16}. Compared to liquid dewetting, solid-state dewetting can be strongly influenced by the anisotropy of the surface energy~\cite{Jiran90,Kim13,Ye11b,Zucker13,Zucker16}, and the mode of mass transport for solid-state dewetting is dominated by surface diffusion~\cite{Srolovitz86a,Mullins57} rather than fluid dynamics. 

In recent years, solid-state dewetting problems have attracted considerate interest, and they have been studied via different mathematical models. From these models, surface diffusion flow and contact line migration have been recognized as the two main kinetic patterns for solid thin films on substrates. The first sharp-interface model for solid-state dewetting was proposed by Srolovitz and Safran~\cite{Srolovitz86a} to investigate the hole growth of small slop profiles under the assumption of isotropic surface energy and cylindrical symmetry. This model was then generalized to both the two-dimensional case~\cite{Wong00} and three-dimensional case~\cite{Du10} in Lagrangian representation. Besides, the "marker particle" method was simultaneously proposed for solving these geometric evolution models. Some other models have been proposed to include the anisotropy, such as the discrete model by Dornel~\cite{Dornel06}, kinetic Monte Carlo models~\cite{Dufay11,Pierre09a} and models via crystalline method~\cite{Kim13,Zucker13}. Moreover, the phase field approaches for solid-state dewetting have also been proposed for both isotropic \cite{Jiang12,Naffouti17} and anisotropic surface energy \cite{Dziwnik15}, which can naturally capture the complex topology change during the evolution. More recently, the two-dimensional solid-state dewetting has been fully studied via sharp interface models~\cite{Wang15,Jiang16,Bao17b,Bao17}. In these models, the interface between the thin film and vapor is assumed to be an open curve with two endpoints (the contact points) attached on the $x$ axis (flat substrate). The open curve evolves via surface diffusion flow while the positions of the two contact points are determined by the relaxed contact angle conditions. Compared to previous models, these model are rigorously derived via the energy variational method, thus account for the full anisotropic free energy of the system and represent a completely mathematical description. The governing equations for the model belongs to fourth-order (for weak anisotropy) or sixth-order (for strong anisotropy) geometric partial differential equations (PDEs) with boundary conditions at the two contact points. These geometric PDEs were then numerically solved by the novel parametric finite element method (PFEM)~\cite{Bao17,Barrett07ISO}, which show great advantages over the previous "marker particle" method. Precisely, "the marker particle" method is an explicit finite difference scheme, thus posing a very severe restriction on time step for the numerical stability. Besides, at each time step, this algorithm requires re-meshing to redistribute the mesh points for the numerical stability. On the contrary, the PFEM overcomes these drawbacks by a semi-implicit mixed form finite element method. By introducing a particular tangential motion for the mesh points, the PFEM has very good properties with respect to the distribution of the mesh points. Although extensive numerical examples for solid-state dewetting have been presented in these papers ~\cite{Wang15,Jiang16,Bao17b,Bao17}, and a generalized Winterbottom construction was proposed to study the equilibrium problem, yet these numerical simulations are restricted to two dimensions and cannot fully demonstrate the complicated geometric evolution during the process of solid-state dewetting.

\begin{figure}[!htp]
\centering
\includegraphics[width=1.0\textwidth]{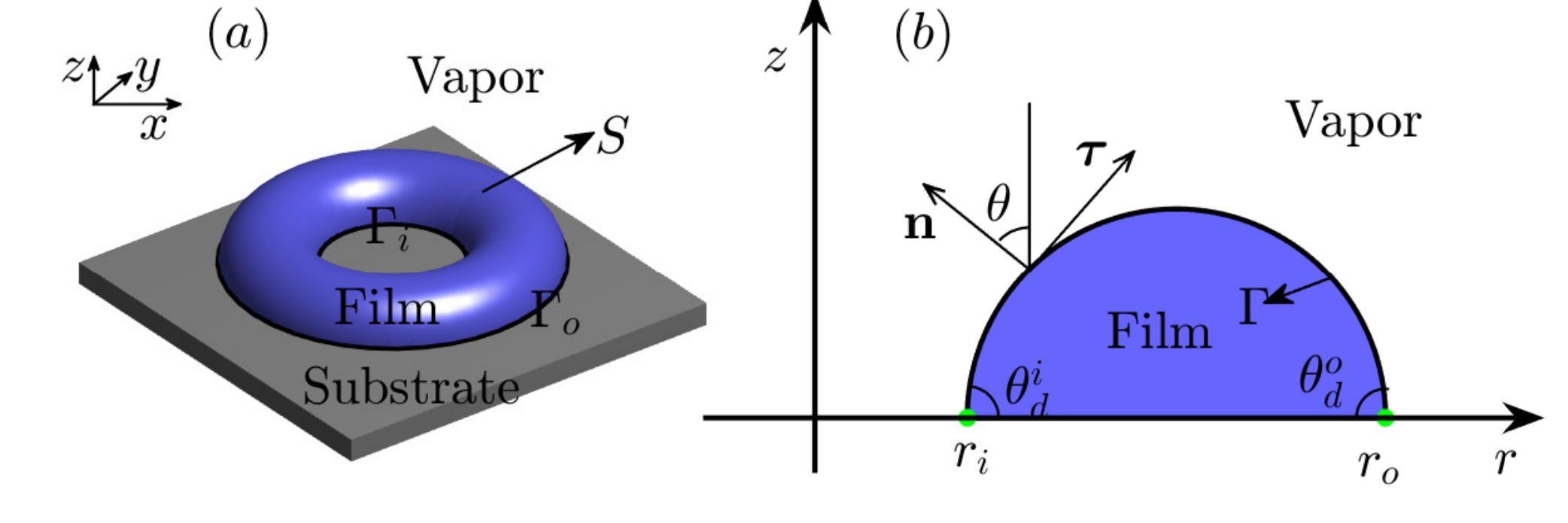}
\caption{A schematic illustration of the solid-state dewetting (a) a toroidal thin film on a flat substrate; (b) The cross-section of an axis-symmetric thin film in the cylindrical coordinate system $(r,z)$. $r_i$ and $r_o$ denotes the radius of inner contact line and outer contact line respectively.}
\label{fig:det}
\end{figure}

In this work, we study the three-dimensional solid-state dewetting problem by imposing the axis-symmetry to both geometry of the thin film and the anisotropy of the surface energy density denoted by $\gamma$. Under these assumptions, the surface evolution problems could collapse to the curve evolution problems by merely considering the cross-section profile of the thin film. As illustrated in Fig.~\ref{fig:det}(a), a cylindrical thin film lies on a flat substrate. The interface that separates the vapor and the thin film is represented by an open surface $S$ with boundaries given by two closed curved $\Gamma_i$ and $\Gamma_o$ on the substrate.
Due to the cylindrical symmetry, the open surface $S$ can be parameterised as follows
\begin{equation}
(s,\varphi)\rightarrow S(s,~\varphi):= \big(r(s)\cos\varphi,~r(s)\sin\varphi,~z(s)\big).
\label{eqn:surface}
\end{equation}
Here in these expressions, $r(s)$ is the radial distance, $\varphi$ is the azimuth angle,
$z(s)$ is the film height, and $s\in [0, L]$ represents the arc length along the axial direction curve (the generatrix, see Fig.~\ref{fig:det}(b)). $S(0,\cdot)$ and $S(L,\cdot)$ represent the inner contact line $\Gamma_i$ and outer contact line $\Gamma_o$, respectively.

The total free energy of the system for solid-state dewetting problems can be written as
\begin{equation}
    W = \iint\limits_S \gamma_{_{\subFV}}(\mathcal{N}) \,dS +\underbrace{(\gamma_{_{\subFS}} - \gamma_{_{\subVS}})A(\Gamma_o\slash\Gamma_i)}_{\rm{Substrate}\;\rm{ energy}}.
    \label{eqn:energy0}
\end{equation}
Here $A(\Gamma_o\slash\Gamma_i)$ denotes the surface area enclosed by the two contact lines, $\gamma_{_{\subFS}}$ and $\gamma_{_{\subVS}}$ are constants and represent the surface energy densities of film/substrate and vapor/substrate respectively. The surface energy density of the film/vapor (surface $S$), given by $\gamma_{_{\subFV}}(\mathcal{N})$, is  assumed to be dependent on the unit out normal vector $\mathcal{N}$ of the surface. The cylindrical symmetry can reduce this dependence to the orientation of curve in the axial direction. Thus we use $\gamma(\theta)=\gamma_{_{\subFV}}(\mathcal{N})$ to denote the surface energy density of film/vapor, with
\begin{eqnarray}
\theta = \arctan\frac{z_s}{r_s};\qquad\gamma(\theta)=\gamma(-\theta),\quad\forall\theta\in[0,\pi];\qquad\gamma(\theta)\in C^2([0,\pi]).
\label{eqn:gammaconditon}
\end{eqnarray}
If we let  $r_o$ and $r_i$ represent the radius of outer contact line and inner contact lines respectively on the substrate, then the total energy can be rewritten as
\begin{equation}
 W = \iint\limits_S  \gamma(\theta) \,dS +\underbrace{(\gamma_{_{\subFS}} - \gamma_{_{\subVS}})(\pi r_o^2-\pi r_i^2)}_{\rm{Substrate}\;\rm{ energy}}.
    \label{eqn:energy}
\end{equation}

Based on the above energy functional, our goals in this paper are (1) to derive the sharp interface-model for solid-state dewetting under the assumption of cylindrical symmetry, (2) to develop a parametric finite element method (PFEM) for solving the derived model based on a good weak formulation, (3) to demonstrate some geometric features during the evolution for solid-state dewetting. We argue that the classical energy variational approaches~\cite{Mullins63,Min06,Ogurtani07} by only considering a normal perturbation of the closed curve or using a graph representation of the interface are not applied to the solid-state dewetting problems. Firstly, the tangential perturbations play an important role in investigating the migration of the contact lines. Secondly, the graph representation of the interface fails to include the case where the isotropic Young angle is big enough. 

 So the rest of the paper is organized as follows. In section $2$, we show rigorously the derivation of sharp interface model via thermodynamic variation and present the equation of equilibrium shape. In section $3$, we present the semi-implicit parametric finite element method for solving the governing geometric PDEs. In section $4$, extensive numerical results will be presented to show the morphological characteristics for solid-state dewetting. In section $5$, we draw a conclusion.

\section{The sharp interface model}
\subsection{Thermodynamic variation}
 Given the parametrisation of the surface $S(s,\varphi)$ in Eq.~\eqref{eqn:surface}, the two tangential vectors of the surface can be calculated directly as follows
\begin{equation}
\mbox{\boldmath{$\mathcal{T}$}}_1=(r_s\cos\varphi,~r_s\sin\varphi,~z_s),\qquad \mbox{\boldmath{$\mathcal{T}$}}_2=(-r\sin\varphi,~r\cos\varphi,~0),
\end{equation}
with subscript $s$ representing taking derivative with respect to $s$. Then the unit outer normal vector of the surface is given by
\begin{equation}
\mbox{\boldmath{$\mathcal{N}$}} = \frac{\mbox{\boldmath{$\mathcal{T}$}}_1\times \mbox{\boldmath{$\mathcal{T}$}}_2}{\abs{\mbox{\boldmath{$\mathcal{T}$}}_1\times \mbox{\boldmath{$\mathcal{T}$}}_2}} = (-z_s\cos\varphi, -z_s\sin\varphi, r_s).
\end{equation}
We consider $s = x_1$ as the first and $\varphi = x_2$ as the second parameter for the surface, then the first
fundamental form is given by
\begin{equation}
 I = E ds^2 + 2F dsd\varphi + G d\varphi^2 ,
 \end{equation}
with $E = r_s^2 + z_s^2 = 1, F = 0, G = r^2.$
Or it can be written in the metric tensor notation as
\begin{equation}
(g_{ij})=\left(
\begin{array}{ll}
1 & 0   \\
0 & r^2
\end{array}
\right),  \ \    g=r^2.
\label{FormI}
\end{equation}
\begin{figure}[!htp]
\centering
\includegraphics[width=1.0\textwidth]{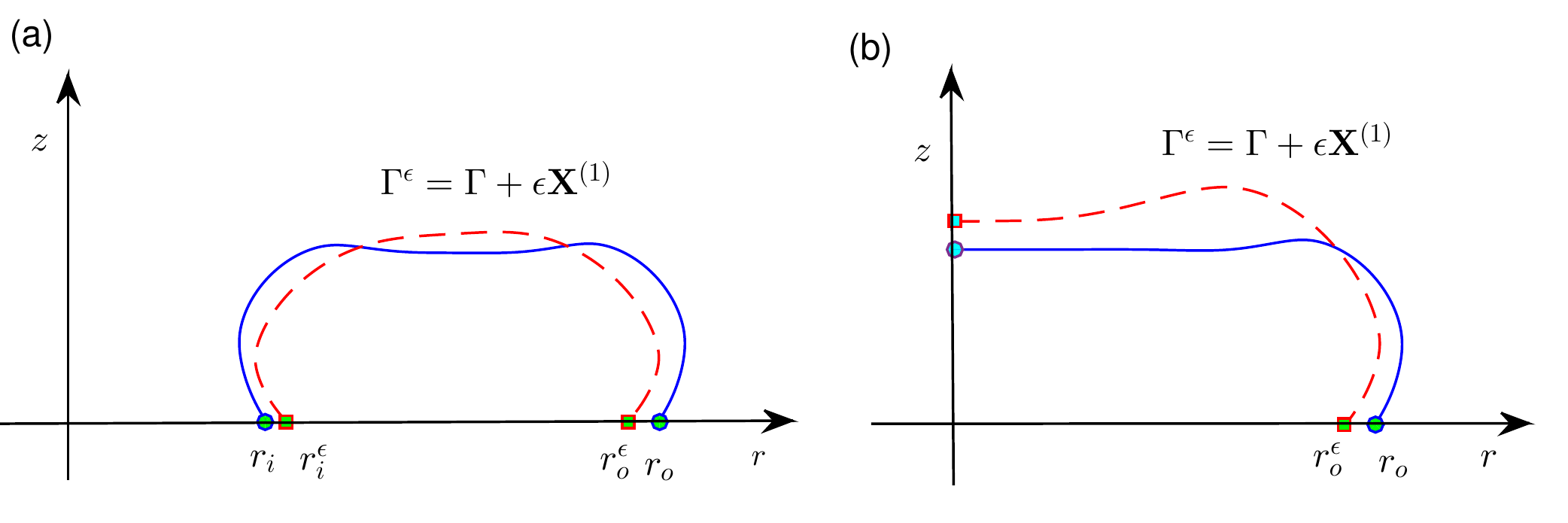}
\caption{A schematic illustration of an infinitesimal perturbation (denoted by the red line) of the curve in axial direction. (a) For toroidal thin film; (b) for island film.}
\label{fig:perp}
\end{figure}

Consider a small axis-symmetric perturbation of the surface $S$ and denote the perturbed surface by $S^\epsilon$:
\begin{equation}
S^\epsilon:=(r^\epsilon(s)\cos\varphi,r^\epsilon(s)\sin\varphi,z^\epsilon(s)).
\end{equation}
 Due to the symmetry, this perturbation can also be viewed as a corresponding perturbation of the curve in the axial direction. As illustrated in Fig.~\ref{fig:perp}, it depicts the perturbations of an initial toroidal thin film (Fig.~\ref{fig:perp}(a)) or island film (Fig.~\ref{fig:perp}(b)), with the red dash line representing the perturbed curve. We can rewrite this perturbation in $(r,z)$ coordinates in a vector form 
 \begin{equation}
\Gamma^\epsilon:=\vec X^\epsilon=(r^\epsilon(s),z^\epsilon(s))=\vec X+\epsilon\vec X^{(1)},
\end{equation}
where $\epsilon$ is a small perturbation parameter and $\vec X^{(1)}=(r^{(1)}(s),z^{(1)}(s))\in (\rm{Lip}[0,L])^2$. Note here if we still take $s$ and $\varphi$ as the two parameters of the surface $S^\epsilon$, the corresponding matrix tensor $g^\epsilon=[(r_s^\epsilon)^2+(z_s^\epsilon)^2](r^\epsilon)^2$.

Based on Eq.~\eqref{eqn:energy}, the total free energy of the new system with the perturbed surface can be written 
\begin{eqnarray}
    W^\epsilon & = & \iint\limits_{S^\epsilon} \gamma(\theta^\epsilon) \,dS^{\epsilon} +\underbrace{(\gamma_{_{\subFS}} - \gamma_{_{\subVS}})(\pi (r_o^\epsilon)^2-\pi (r_i^\epsilon)^2)}_{\rm{Substrate\quad energy}}\nonumber\\
      & = & \int_0^{2\pi}\int_0^L\,\gamma(\theta^\epsilon)\sqrt{g^\epsilon}\,ds\,d\phi+(\gamma_{_{\subFS}} - \gamma_{_{\subVS}}) \pi [r^\epsilon(L)^2 - r^\epsilon(0)^2]\nonumber\\
      & = & 2\pi\int_0^L\,\gamma(\theta^\epsilon)|\vec X_s^\epsilon|r^\epsilon\,ds+ (\gamma_{_{\subFS}} - \gamma_{_{\subVS}})\pi [r^\epsilon(L)^2 - r^\epsilon(0)^2].
      \label{eqn:penergy}
\end{eqnarray}

The unit tangential and outer normal vector of the curve $\Gamma$ are given as $\boldsymbol{\tau}=\partial_s\vec X$ and $\vec n=-(\partial_s\vec X)^\perp$, where $\perp$ denotes a clockwise rotation of 90 degrees of a vector. We expand the following terms at $\epsilon=0$ and obtain
\begin{subequations}
\begin{align}
\label{eqn:expansion1}
&r^\epsilon  =  r + r^{(1)}\epsilon+O(\epsilon^2),\\
\label{eqn:expansion2}
&|\vec {X_s^\epsilon}|  =  1 + (\boldsymbol{\tau}\cdot\vec X_s^{(1)})\epsilon + O(\epsilon^2),\\
&\gamma(\theta^\epsilon)  =  \gamma(\theta) + \gamma'(\theta)(\vec n\cdot\vec X_s^{(1)})\epsilon + O(\epsilon^2).
\label{eqn:expansion3}
\end{align}
\end{subequations}
Now in order to calculate the first variation of the total energy of the system, we can write the total energy as 
\begin{equation}
W^\epsilon=W+ W^{(1)} \epsilon+O(\epsilon^2).
\end{equation}
Here $W^{(1)}$ is the coefficient of the order $\epsilon$ term, which can be obtained directly by substituting Eq.~\eqref{eqn:expansion1}-\eqref{eqn:expansion3} into Eq.~\eqref{eqn:penergy}
\begin{eqnarray} 
W^{(1)}&=&2\pi\int_0^L\Bigl[\gamma'(\theta)(\vec n\cdot\vec X_s^{(1)})r+\gamma(\theta)(\boldsymbol{\tau}\cdot\vec X_s^{(1)})r
+\gamma(\theta)r^{(1)}\Bigr]\,ds\nonumber\\&&+\;2\pi(\gamma_{_{\subFS}} - \gamma_{_{\subVS}})\big[r(L)r^{(1)}(L) - r(0)r^{(1)}(0)\big].
\end{eqnarray}
Integration by parts for the above equation, we obtain
\begin{eqnarray}
W^{(1)}&=&2\pi\int_0^L\Bigl[-(\gamma(\theta)\boldsymbol{\tau}+\gamma'(\theta)\vec n)_s\cdot\vec X^{(1)} r-(\gamma(\theta)\boldsymbol{\tau}+\gamma'(\theta)\vec n)\cdot\vec X^{(1)}r_s+\gamma(\theta)r^{(1)}\Bigr]\,ds\nonumber\\
&&+\;2\pi\Bigl([\gamma(\theta)\boldsymbol{\tau}+\gamma'(\theta)\vec n]\cdot\vec X^{(1)}r\Bigr)\Big|_{s=0}^{s=L}\nonumber\\&&+ \;2\pi (\gamma_{_{\subFS}} - \gamma_{_{\subVS}})\big[r(L)r^{(1)}(L) - r(0)r^{(1)}(0)\big].
\label{eqn:Boundaryterms}
\end{eqnarray}
Let $\kappa=-\vec X_{ss}\cdot\vec n$ be the curvature of curve $\Gamma$. Moreover, for simplicity, we assume that the curve $\Gamma$ has two contact points $r_i,r_o$ (see Fig.~\ref{fig:perp}(a)). Denote $\theta_d^i,\,\theta_d^o$ as the inner and outer contact angle of curve $\Gamma$ respectively, we have the following expressions:
\begin{eqnarray*}
&&\vec n_s=\kappa\boldsymbol{\tau},\quad\vec n\Big|_{s=0}=(-\sin\theta_d^i,\cos\theta_d^i),\quad\vec n\Big|_{s=L}=(\sin\theta_d^o,\cos\theta_d^o),\quad\theta_s=-\kappa,\\
&&\boldsymbol{\tau}_s=-\kappa\vec n,\quad\boldsymbol{\tau}\Big|_{s=0}=(\cos\theta_d^i,\sin\theta_d^i),\quad\boldsymbol{\tau}\Big|_{s=L}=(\cos\theta_d^o,-\sin\theta_d^o),\quad\boldsymbol{\tau}\cdot\vec X^{(1)}r_s-r^{(1)}=z_s\vec X^{(1)}\cdot\vec n.
\end{eqnarray*}
In addition, we requires that the two contact lines are always on the substrate after the small cylindrical perturbation, \begin{equation*}
\vec X^{(1)}\Big|_{s=0}=(r^{(1)}(0),0),\quad\vec X^{(1)}\Big|_{s=L}=(r^{(1)}(L),0).
\end{equation*}
Then we immediately obtain
\begin{eqnarray}
W^{(1)}&=&2\pi\int_0^L\Bigl[(\gamma(\theta)+\gamma''(\theta))\kappa(\vec n\cdot\vec X^{(1)})r-(\gamma(\theta)z_s+\gamma'(\theta)r_s)(\vec X^{(1)}\cdot\vec n)\Bigr]\,ds\nonumber\\
&&+\;2\pi r(L)r^{(1)}(L)\Bigl[\gamma(\theta_d^o)\cos\theta_d^o-\gamma'(\theta_d^o)\sin\theta_d^o+(\gamma_{_{\subFS}} - \gamma_{_{\subVS}})\Bigr]\nonumber\\
&&-2\;\pi r(0)r^{(1)}(0)\Bigl[\gamma(\theta_d^i)\cos\theta_d^i-\gamma'(\theta_d^i)\sin\theta_d^i+(\gamma_{_{\subFS}} - \gamma_{_{\subVS}})\Bigr]\nonumber\\
&=&\iint\limits_S\Bigl[(\gamma(\theta)+\gamma''(\theta))\kappa^0-\frac{\gamma(\theta)z_s+\gamma'(\theta)r_s}{r}\Bigr]\vec n\cdot\vec X^{(1)}\,dS\nonumber\\
&&+\;\int_{\Gamma_o}\Bigl[\gamma(\theta_d^o)\cos\theta_d^o-\gamma'(\theta_d^o)\sin\theta_d^o+(\gamma_{_{\subFS}} - \gamma_{_{\subVS}})\Bigr]r^{(1)}(L)\,d\Gamma\nonumber\\
&&-\;\int_{\Gamma_i}\Bigl[\gamma(\theta_d^i)\cos\theta_d^i-\gamma'(\theta_d^i)\sin\theta_d^i+(\gamma_{_{\subFS}} - \gamma_{_{\subVS}})\Bigr]r^{(1)}(0)\,d\Gamma.
\label{eqn:varenergy}
\end{eqnarray}
From the above equations, we immediately obtain the variation of the total energy with respect to the surface and the two contact lines as
\begin{eqnarray}
&&\frac{\delta W}{\delta S}=(\gamma(\theta)+\gamma''(\theta)\kappa-\frac{\gamma(\theta)z_s+\gamma'(\theta)r_s}{r},\\
&&\frac{\delta W}{\delta \Gamma_o}=\gamma(\theta_d^o)\cos\theta_d^o-\gamma'(\theta_d^o)\sin\theta_d^o+(\gamma_{_{\subFS}} - \gamma_{_{\subVS}}),\\
&&\frac{\delta W}{\delta \Gamma_i}=-\Big[\gamma(\theta_d^i)\cos\theta_d^i-\gamma'(\theta_d^i)\sin\theta_d^i+(\gamma_{_{\subFS}} - \gamma_{_{\subVS}})\Big].
\end{eqnarray}

Note here the variation is calculated for the case when the thin film has two contact lines. For island film, as depicted in Fig.~\ref{fig:perp}(b), there is no inner contact line. Thus after the perturbation, we require at the boundary the following equations should be satisfied for the island film
\begin{equation}
\vec X^{(1)}\Big|_{s=0}=(0,z^{(1)}(0)),\qquad \vec X^{(1)}\Big|_{s=L}=(r^{(1)}(L),0).
\end{equation}
When there doesn't exist the inner contact line, integration by parts in Eq.~\eqref{eqn:Boundaryterms} will not produce boundary terms at $s=0$ ($r=0$).

\subsection{The model and its properities}
From the anisotropic Gibbs-Thomson relation \cite{Sutton95}, the chemical potential can be defined as 
\begin{equation}
 \mu=\Omega_0\frac{\delta W}{\delta S}=\Omega_0[(\gamma(\theta)+\gamma''(\theta))\kappa-\frac{\gamma(\theta)z_s+\gamma'(\theta)r_s}{r}],
 \end{equation}
with $\Omega_0$ representing the atomic volume. Subsequently the normal velocity of the surface is given by surface diffusion \cite{Mullins57,Cahn74}
\begin{equation}
v_n=\frac{D_s\nu\Omega_0}{k_B\,T_e}\nabla_{_S}^2\mu.
\end{equation}
In these expressions, $D_s$ is the surface diffusivity, $k_B\,T_e$ is the thermal energy, $\nu$ is the number of diffusing atoms per unit area, $\nabla_{_S}$ is the surface gradient.
The two contacts lines $\Gamma_i,\,\Gamma_o$ move on the substrate, and the velocities $v_c^i,\,v_c^o$ are given by the time-dependent Ginzburg-Landau kinetic equations \cite{Wang15,Jiang16},
\begin{eqnarray}
&&v_c^o=-\eta\frac{\delta W}{\delta\Gamma_o}=-\eta\Big[\gamma(\theta_d^o)\cos\theta_d^o-\gamma'(\theta_d^o)\sin\theta_d^o+(\gamma_{_{\subFS}} - \gamma_{_{\subVS}})\Big],\\
&&v_c^i=-\eta\frac{\delta W}{\delta\Gamma_i}=\eta\Big[\gamma(\theta_d^i)\cos\theta_d^i-\gamma'(\theta_d^i)\sin\theta_d^i+(\gamma_{_{\subFS}} - \gamma_{_{\subVS}})\Big],
\end{eqnarray}
with $\eta\in(0,+\infty)$ denoting the contact line mobility.

Now take the characteristic length scale and characteristic surface energy scale as $L$ and $\gamma_0$ respectively, by choosing the time scale as $\frac{L^4}{B\gamma_0}$ with $B=\frac{D_s\nu\Omega_0^2}{k_B\,T_e}$, and the contact line mobility is scaled by $\frac{B}{L^3}$. Note that  $\nabla_{_S}^2 \mu = \frac{1}{\sqrt{g}}\partial_i(\sqrt{g}g^{ij}\partial_j \mu)  = \frac{1}{r}\left(r \mu_s\right)_s$. Thus if we let $\Gamma(t)=\vec X(s,t)$ represent the moving curve in the axial direction, the sharp interface model for solid-state dewetting with anisotropic surface energy in three dimensions with cylindrical symmetry can be described in the following dimensionless form
\begin{subequations}
\begin{align}
\label{eqn:pde1}
&\partial_t\vec X=\frac{1}{r}\partial_s(r\partial_s\mu)\,\vec n,\quad 0<s<L(t),\quad t>0,\\
\label{eqn:pde2}
&\mu=[\gamma(\theta)+\gamma''(\theta)]\kappa-\frac{\gamma(\theta)\partial_s z+\gamma'(\theta)\partial_s r}{r},\\
&\kappa=-(\partial_{ss}\vec X)\cdot\vec n;
\label{eqn:pde3}
\end{align}
\end{subequations}
where $L:=L(t)$ represents the total arc length of the curve $\Gamma(t)$, all the variables are in dimensionless form and we still use the same notation for brevity. The initial condition is given as
\begin{equation}\label{init}
\vec{X}(s,0):=\vec{X}_0(s)=(r(s,0),z(s,0))=(r_0(s),z_0(s)), \qquad 0\le s\le L_0:=L(0).
\end{equation}
The above governing equations are together with the following boundary conditions
\begin{itemize}
\item[(i)] contact line condition
\begin{equation}
z(L, t) = 0,\qquad
\begin{cases}z(0,t)=0,&\;\rm{if}\; r(0,t)>0,\\
\partial_sz(0,t)=0,&\;\rm{otherwise},
\end{cases}
\quad t\geq 0;\label{eqn:pdebc1}
\end{equation}
\item[(ii)] relaxed (or dissipative) contact angle condition
\begin{equation}
\label{eqn:pdebc2}
\partial_t r(L, t)= -\eta f(\theta_d^o;\sigma),\quad 
\begin{cases}
\partial_t r(0,t)= \eta f(\theta_d^i;\sigma),&\;\rm{if}\;r(0,t)>0,\\
r(0,t)=0,&\;\rm{otherwise},
\end{cases}
\quad t\geq0;
\end{equation}
\item[(iii)] zero-mass flux condition
\begin{equation}
\partial_s \mu(0, t)=0,\quad  \partial_s\mu(L, t) = 0,\quad t\geq 0;
\label{eqn:pdebc3}
\end{equation}
\end{itemize}
where $\theta_d^i,\theta_d^o$ are the (dynamic) contact angles at the inner contact line and outer contact line,
respectively, $0 <\eta<\infty$ denotes the dimensionless contact line mobility, and $f(\theta;\sigma)$ is defined as
\begin{equation}
f(\theta;\sigma) = \gamma(\theta)\cos\theta-\gamma'(\theta)\sin\theta-\sigma,\quad\theta\in[-\pi,\pi],\quad\sigma=\frac{ \gamma_{_{\subVS}}-\gamma_{_{\subFS}}}{\gamma_0}.
\end{equation}
 In the case when the thin film does not have an inner contact line, $r(0,t)$ is fixed at $0$ and the symmetry for the island also require the Neumann boundary condition for $z$ at $s=0$, see Eq.~\eqref{eqn:pdebc1}. 

In the following, we will show that the total volume of the thin film (the volume enclosed by the surface $S(t)$ and the substrate) is conserved and the total energy of the system is dissipative. We introduce a new parameter $\rho\in I=[0,1]$ to parameterize the evolution curves $\Gamma(t)$ such that $\Gamma(t)$ is a family of open curves, where $t\in[0,T]$, 
\begin{equation}
\Gamma(t)=\vec X(\rho, t): I\times [0,T]\rightarrow \mathbb{R}^2.
\end{equation}
It should be noted that the relationship between the parameter $\rho$ and the arc length $s$ can be given as $s(\rho,t)=\int_0^\rho |\partial_\rho\vec{X}|\;d\rho$, and then we can obtain that $\partial_\rho s=|\partial_\rho\vec{X}|$.

\begin{prop}[Mass conservation]\label{prop:massconservation} Suppose $\vec X(\rho,t),\mu(\rho,t)$ and $\kappa(\rho,t)$ satisfy the PDEs~\eqref{eqn:pde1}-\eqref{eqn:pde2} with boundary conditions \eqref{eqn:pdebc1}-\eqref{eqn:pdebc3}. Let $V(t)$ denote the volume of the thin film, we then have:
\begin{equation*}
V(t)=V(0),\qquad\forall 0\le t\le T.
\end{equation*}
\begin{proof}
We know we can write $V(t)$ as
\begin{equation*}
V(t)=\int_0^{2\pi}\int_{0}^{L(t)}r z r_s\;ds\;d\varphi.
\end{equation*}
Take derivative with respect to $t$ for $V(t)$ we obtain directly
\begin{eqnarray*}
\frac{d}{dt}V(t)&=&2\pi\frac{d}{dt}\int_{0}^1 rr_\rho z\;d\rho=2\pi\int_0^1(r_t r_\rho z+r r_\rho z_t)\;d\rho+2\pi\int_0^1 r r_{\rho,t} z\;d\rho\\
&=&2\pi\int_0^1(r_\rho z r_t+r r_\rho z_t)\;d\rho-2\pi\int_0^1(r z)_\rho r_t\;d\rho+2\pi\Bigl(r z r_t\Bigr)\Big|_{\rho=0}^{\rho=1}
\end{eqnarray*}
By noting Eqs.~\eqref{eqn:pdebc1},\eqref{eqn:pdebc2}, the boundary terms vanished. Thus above equation can be simplified as flows
\begin{eqnarray*}
\frac{d}{dt}V(t)&=&2\pi\int_0^1(r r_\rho z_t-r z_\rho r_t)\;d\rho=2\pi\int_0^{1} r\partial_t\vec X\cdot(-\vec X_\rho)^{\perp}\;d\rho\\
&=&2\pi\int_0^{L(t)} r\partial_t\vec X\cdot\vec n\;ds=2\pi\int_0^{L(t)} \partial_s(r\partial_s\mu)\;ds=0.
\end{eqnarray*}
Here we have using the Eq.~\eqref{eqn:pde1} to replace $\partial_t\vec X\cdot\vec n$ by $\frac{1}{r}\partial_s(r\partial_s\mu)$.
 Thus the mass is conserved during the time evolution.
\end{proof}
\end{prop}

\begin{prop}[Energy dissipation]\label{prop: energydissipation}Suppose $\vec X(\rho,t),\mu(\rho,t)$ and $\kappa(\rho,t)$ satisfy the PDEs~\eqref{eqn:pde1}-\eqref{eqn:pde2} with boundary conditions \eqref{eqn:pdebc1}-\eqref{eqn:pdebc3}. Let $W(t)$ denote the energy of the whole system, we then have:
\begin{equation*}
W(0)\ge W(t_1)\ge W(t_2),\qquad\forall\;0\le t_1\le t_2.
\end{equation*}
\begin{proof}
In order to obtain the time derivative of the total surface energy, we can make use of the first variation of the total surface energy we calculated previously.  
By replacing the perturbation parameter $\varepsilon$ with the time $t$, and note that first order perturbation $X^{(1)}$ will then be replaced by $\partial_t\vec X$, we can easily obtain that the time derivative of the total energy according to Eq.~\eqref{eqn:varenergy}
\begin{eqnarray}
\frac{d}{dt}W(t)&=&2\pi\int_0^{L(t)}r\vec n\cdot\partial_t\vec X\big[(\gamma(\theta)+\gamma''(\theta))\kappa-\frac{z_s\gamma(\theta)+r_s\gamma'(\theta)}{r}\big]\;ds\nonumber\\
&&+\int_{\Gamma_o}\Bigl[\gamma(\theta_d^o)\cos\theta_d^o-\gamma'(\theta_d^o)\sin\theta_d^o-\sigma)\Bigr]\frac{dr_o}{dt}\,d\Gamma\nonumber\\
&&-\int_{\Gamma_i}\Bigl[\gamma(\theta_d^i)\cos\theta_d^i-\gamma'(\theta_d^i)\sin\theta_d^i-\sigma)\Bigr]\frac{dr_i}{dt}\,d\Gamma.
\end{eqnarray}
Combining with Eq.~\eqref{eqn:pde1},~\eqref{eqn:pde2}, we have
\begin{eqnarray*}
\frac{d}{dt}W(t)
&=&2\pi\int_o^{L(t)}(r\mu_s)_s\mu\;ds-\frac{2\pi}{\eta}\big[r_{o}(\frac{dr_{_o}}{dt})^2+r_{_i}(\frac{dr_{_i}}{dt})^2\big]\\
&=&-2\pi\int_0^{L(t)}r(\mu_s)^2\;ds-\frac{2\pi}{\eta}\big[r_{o}(\frac{dr_{_o}}{dt})^2+r_{_i}(\frac{dr_{_i}}{dt})^2\big]\le 0,
\end{eqnarray*}
which immediately implies the energy dissipation. 
\end{proof}
\end{prop}
In the proof of the energy dissipation, we only show the case when the island forms two contact lines. When there is no inner contact line, the energy dissipation is still satisfied as the boundary term at $\rho=0$ will no longer exist. 

\subsection{Equilibrium shape}
The equilibrium shape for the solid-state dewetting problem is a minimization of the total surface energy under the constraint such that the total volume of the thin film is fixed at constant $C>0$. 
The problem can be
mathematically expressed in the following
  \begin{equation}
  \min_{\Omega}\big(\int_0^{2\pi}\int_0^L \gamma(\theta) \,ds\,d\phi - \sigma(\pi r_o^2-\pi r_i^2)\big),
 \quad \rm{s.t.}\quad \abs{\Omega}=C.
  \end{equation}
By introducing the Lagrange multiplier $\lambda$, we can define the Lagrange function
\begin{eqnarray}
L(\Gamma,\lambda) & = & W - \lambda \big(|\Omega|-C\big)\nonumber\\
& = & \int_0^{2\pi}\int_0^L \gamma(\theta) \,ds\,d\phi - \sigma(\pi r_o^2-\pi r_i^2) - \lambda \big(\int_0^{2\pi}\int_{0}^{L}r z r_s\;ds\;d\varphi - C\big).
\end{eqnarray}

Based on the first variation of the dimensionless total surface energy and variation of the total volume, we know that the equation for the equilibrium shape can be obtained by vanishing the first variation of the Lagrange function
\begin{equation}
0=\delta L = \iint_S (\mu - \lambda) \vec n\cdot\vec X^{(1)}\;dS + \sigma\int_{\Gamma_o}f(\theta_d^o,\sigma)r^{(1)}(L)\,d\Gamma-\sigma\int_{\Gamma_i}f(\theta_d^i,\sigma)r^{(1)}(0)\,d\Gamma,
\end{equation}
Assume the curve of the equilibrium shape in the axis direction is given by $\Gamma_e:=(r(s),z(s)),\;s\in[0,L]$, then in order to ensure the first variation of the surface energy to be zero, it is equivalent to the following two conditions
\begin{subequations}
\begin{align}
&\mu(s):=[\gamma(\theta)+\gamma^{\prime\prime}(\theta)]\kappa - \frac{\gamma(\theta)\partial_s z+\gamma^\prime(\theta)\partial_s r}{r}\equiv \lambda,\quad a.e.\;s\in[0,L],\\
&r(0)=0,\quad f(\theta_o^c;\sigma)=\gamma(\theta_c^o)\cos\theta-\gamma^\prime(\theta_c^o)\sin\theta_c^o-\sigma=0.
\end{align}
\end{subequations}
Following the Winterbottom construction \cite{Winterbottom67,Bao17b}, we know the equation for the equilibrium shape for solid-state dewetting under the cylindrical symmetry can be given as follows up to a scaling
\begin{equation}
\begin{cases}
r(\theta) = \gamma(\theta)\sin\theta + \gamma'(\theta)\cos\theta,\\
z(\theta) = \gamma(\theta)\cos\theta - \gamma'(\theta)\sin\theta-\sigma,
\end{cases}
\qquad\theta\in[0,\theta_c^o].
\end{equation}
When $\theta=0$, in view of Eq.~\eqref{eqn:gammaconditon}, we obtain $r(0)=0$, this implies that no inner contact line exists in the equilibrium shape. $\theta_c^o$ is the static contact angle for the outer contact line, and it satisfies $f(\theta_c^o;\sigma)=0$, which immediately gives $z(\theta_c^o)=0$.

\section{Parametric finite element method}
\subsection{Variational formulation}
To formulate the variational formulation for the geometric PDEs: Eqs.~\eqref{eqn:pde1},~\eqref{eqn:pde2},~\eqref{eqn:pde3} with boundary conditions Eqs.~\eqref{eqn:pdebc1},~\eqref{eqn:pdebc2},~\eqref{eqn:pdebc3}, we need to introduce some functional spaces with respect to the evolution curve $\Gamma(t)$. First, we can define the space on the domain $I$ as 
\begin{equation*}
L^2(I)=\{f: I\rightarrow \mathbb{R}, \;\text{and} \int_{\Gamma(t)}|f(s)|^2 ds
=\int_I |f(s(\rho,t))|^2 \partial_\rho s d\rho <+\infty \},
\end{equation*}
equipped with the $L^2$ inner product $\big<f,g\big>_{\Gamma}:=\int_{\Gamma}f\cdot g\,ds$ for any scalar (or vector) functions $f,g$. Moreover, we define the the following two special functional spaces:
\begin{eqnarray}
&&H_{a,b}^{(r)}(I)=\{\phi\in H^1(I):\;~\phi(0)=a;\;\phi(1)=b\},\\
&&H_{a,b}^{(z)}(I)=\{\phi\in H^1(I):\;~\phi(1)=0;\;\text{if}\;~a>0,~~\phi(0)=0\},
\end{eqnarray}
with $a$, $b$ given as two parameters related to the radius of the two moving contact lines. It is obvious that from the above definition, we have $H_{0,0}^{(r)}(I)=H_0^1(I)$.

Let $\widetilde{\gamma}(\theta)=\gamma(\theta)+\gamma''(\theta)$, $\vec b(\theta)=(\gamma'(\theta),\gamma(\theta))$, then the variational formulation for solid-state dewetting in three dimensions with cylindrical symmetry can be stated as follows: given $\Gamma(0)=\vec X(\rho,0)=(r(\rho,0),z(\rho,0))$, for any time $t\in[0,T]$, we want to find $\Gamma(t)=\vec X(\rho,t)\in H_{a,b}^{(r)}(I)\times H_{a,b}^{(z)}(I)$ with $r-$coordinates positions of two contact lines $0\leq r(0,t)=a\le b=r(1,t)$ and $\mu(\rho,t)\in H^1(I)$, $\kappa(\rho,t)\in H^1(I)$ such that:
\begin{subequations}
\begin{align}
&\big<\partial_t\vec X\cdot\vec n,~\phi\;r\big>_\Gamma+\big<\partial_s\phi,~\partial_s\mu\;r\big>_\Gamma=0,\qquad\forall\phi\in H^1(I),\label{eqn:vf1}\\
&\big<\mu,~\varphi\;r\big>_\Gamma -\big<\widetilde{\gamma}(\theta)\kappa,~\varphi\;r\big>_\Gamma+\big<\vec b(\theta)\cdot\boldsymbol{\tau},~\varphi\big>_\Gamma=0,\qquad\forall\varphi\in H^1(I),\label{eqn:vf2}\\
&\big<\kappa,~\vec n\cdot\boldsymbol{\omega}\big>_\Gamma -\big<\partial_s\vec X,~\partial_s\boldsymbol{\omega}\big>_\Gamma=0,\qquad\forall\boldsymbol{\omega}\in H_0^1(I)\times H_{a,b}^{(z)}(I).\label{eqn:vf3}
\end{align}
\end{subequations}
Eq.~\eqref{eqn:vf2} is obtained directly by multiplying $r\varphi$ for Eq.~\eqref{eqn:pde2} and then integrating over $\Gamma$. For Eq.~\eqref{eqn:pde1}, firstly we multiply $r\vec n$ to  cancel the fraction term, and then multiply the test function $\phi$. By noting the boundary condition Eq.~\eqref{eqn:pdebc3}, integrating by parts over $\Gamma$ immediately gives Eq.~\eqref{eqn:vf1}. Moreover, by re-formulating Eq.~\eqref{eqn:pde3} as $\kappa\;\vec n=-\partial_{ss}\vec X$, and then multiplying the vector valued test function $\boldsymbol{\omega}$, integrating by parts, Eq.~\eqref{eqn:vf3} can be derived in view of the boundary condition Eq.~\eqref{eqn:pdebc1}. For the cases $a=0$ or $a>0$, the variational formulation are different as the functional space for the test function $\boldsymbol{\omega}$ depends on $a$.

\subsection{Fully discrete scheme}
To present parametric finite element method (PFEM) for the variational problem \eqref{eqn:vf1},~\eqref{eqn:vf2},~\eqref{eqn:vf3}, we first discrete the time as $0=t_0<t_1<t_2<\ldots t_M$ with the time step $\tau_m=t_{m+1}-t_m$. Also a uniform partition of the domain $I$ is given as $I=[0,1]=\bigcup_{j=1}^{N}I_j$ with $h=\frac{1}{N}$ denoting the mesh size, and $I_j=[\rho_{j-1},\rho_{j}]$.
Introduce the following finite dimensional approximations to $H^1(I)$, $H_{a,b}^{(z)}(I)$ and $H_{a,b}^{(z)}(I)$ 
\begin{eqnarray}
\label{eqn:FEMspace1}
&&V^h:=\{u\in C(I):\;u\mid_{I_{j}}\in P_1,\quad  j=1,2,\ldots,N\}\subset H^1(I),\\
\label{eqn:FEMspace2}
&&\mathcal{V}^{h,(r)}_{a,b}:=\{u \in V^h:\;u(0)=a;~u(1)=b\}\subset H^{(r)}_{a,b}(I),\\
&&\mathcal{V}^{h,(z)}_{a,b}:=\{u \in V^h:\;\text{ if }a>0,~u(0)=0;~~u(1)=0\}\subset H^{(z)}_{a,b}(I),
\end{eqnarray}
where $P_1$ denotes all polynomials with degrees at most $1$, $a$ and $b$ are two given constants.

Let $\Gamma^{m}:=\vec{X}^{m}=(r^m,z^m)$, $\vec{n}^m$, $\mu^m$, $\theta^m$ and $\kappa^m$ be the numerical approximations of the moving curve $\Gamma(t_m):=\vec{X}(\cdot,t_m)$, the normal vector $\vec{n}$, the chemical potential $\mu$, the angle $\theta$ and the curvature
$\kappa$ at time $t_{m}$, respectively. The polygonal curve $\Gamma^m$ can be written as
\begin{equation}
\Gamma^m = \bigcup_{j=1}^{N}\bar{h}_j^m, \quad\{h_j^m\}_{j=1}^{N} \;\text{are connected line segments of the curve}\;\Gamma^m.
\end{equation}
Here these line segments are order in the clockwise direction. By noting the relationship between the parameter $\rho$ and $s$, we can obtain the the following equation for a smooth function $\phi(s)$
\begin{equation}
\partial_s\phi=\frac{\partial_\rho\phi}{\partial_\rho s}=\frac{\partial_\rho\phi}{|\partial_\rho\vec X^m|}.
\end{equation}
Thus, the unit tangential vector $\boldsymbol{\tau}^m$ is a step function with discontinuities on the endpoints of each line segment $h_j^m$. It can be numerically calculated as
\begin{equation}
\boldsymbol{\tau}^m\Big|_{h_j^m}= \partial_s\vec X^m\Big|_{h_j^m}=\frac{\vec X^m(\rho_j)-\vec X^m(\rho_{j-1})}{|\vec X^m(\rho_j)-\vec X^m(\rho_{j-1})|},\quad\forall 1\leq j\leq N.
\end{equation}
Consequently, the normal vector $\vec n^m$ and angle $\theta^m$ can be obtained directly
\begin{equation}
\vec n^m\Big|_{h_j^m} = -{\boldsymbol{\tau}^m}^\perp\Big|_{h_j^m},\qquad \theta^m\Big|_{h_j^m}=\arctan\frac{z^m(\rho_j)-z^m(\rho_{j-1})}{r^m(\rho_j)-r^m(\rho_{j-1})}.
\end{equation}
where $\perp$ means a $90$ degree's clockwise rotation of a vector. Moreover, given two scalar or vector functions $f,g$, which is defined over $\Gamma^m$, define the following discrete norms to approximate the integration over $\Gamma^m$ 
\begin{equation}
\big<f,~g\big>_{\Gamma^m} = \frac{1}{6}\sum_{j=1}^{N}|h_j^m|\Bigl[f((\rho_{j-1})^+)\cdot g((\rho_{j-1})^+)+ 4f(\rho_{j-\frac{1}{2}})\cdot g(\rho_{j-\frac{1}{2}}) + f((\rho_{j})^-)\cdot g((\rho_{j})^-)\Bigr].
\end{equation}
Here $f(\rho^-)$ and $f(\rho^+)$ are one-sided limit and can be defined respectively as
\begin{equation}
f(\rho^{\pm})=\lim_{\epsilon\rightarrow 0^+}f(\rho\pm\epsilon).
\end{equation}
Now take $\Gamma^0=\vec{X}^0\in \mathcal{V}^{h,(r)}_{r_0(0),r_0(L_0)} \times \mathcal{V}^{h,(z)}_{r_0(0),r_0(L_0)}$
such that $\vec{X}^0(\rho_j)=\vec{X}_0(s_j^0)$ with $s_j^0=jL_0/N=L_0\rho_j$ for $j=0,1,\ldots,N$ and obtain
$\vec n^0,\,\theta^0$ via the initial data \eqref{init},
then  a semi-implicit {\sl parametric finite element method (PFEM)} for the variational problem
\eqref{eqn:vf1}-\eqref{eqn:vf3} can be given as: for $m\ge0$, first update the radius of the two contact line positions via
the relaxed contact angle conditions~\eqref{eqn:pdebc2} by using the forward Euler method or simply restrict $r_{i}(t_{m+1})=0$ if there is no inner contact line,
and then
find $\Gamma^{m+1}=\vec{X}^{m+1}\in \mathcal{V}^{h,(r)}_{a,b}\times
\mathcal{V}^{h,(z)}_{a,b}$ with the $r$-coordinate positions of the
moving contact lines $a:=r_{i}(t_{m+1})\leq b:=r_{o}(t_{m+1})$, $\mu^{m+1}\in V^h$ and
$\kappa^{m+1}\in V^h$ such that
\begin{subequations}
\begin{align}
\label{eqn:fd1}
&\big<\vec X^{m+1}-\vec X^m\cdot\vec n^m,\;\phi_h\;r^m\big>_{\Gamma^m}+\tau_m\big<\partial_s\mu^{m+1},\;\partial_s\phi_h\;r^m\big>_{\Gamma^m}=0,\quad\forall\phi_h\in V^h,\\
\label{eqn:fd2}
&\big<\mu^{m+1},\;\varphi_h\;r^m\big>_{\Gamma^m} -\big<\tilde{\gamma}(\theta^m)\kappa^{m+1},\;\varphi_h\;r^m\big>_{\Gamma^m}+\big<\vec b(\theta^m)\cdot\boldsymbol{\tau}^m,\;\varphi_h\big>_{\Gamma^m}=0,\quad\forall\varphi_h\in V^h,\\
&\big<\kappa^{m+1},\;\vec n^m\cdot\boldsymbol{\omega}_h\big>_{\Gamma^m} -\big<\partial_s\vec X^{m+1},\;\partial_s\boldsymbol{\omega}_h\big>_{\Gamma^m}=0,\quad\forall\boldsymbol{\omega}_h\in \mathcal{V}^{h,(r)}_{0,0}\times \mathcal{V}^{h,(z)}_{a,b}.
\label{eqn:fd3}
\end{align}
\end{subequations}

In the above semi-implicit PFEM, we can observe that the nonlinear terms $\tilde{\gamma}(\theta),~b(\theta)$ are valued explicitly. Moreover, the numerical integration is calculated over $\Gamma^m$ instead of $\Gamma^{m+1}$, and the unit normal and tangential vector, the $r$ coordinates are also valued on $\Gamma^m$. These particular treatments could eventually lead the discrete scheme to be a linear system with the sparse matrix. During our practical computation, the linear system can be efficiently solved via the sparse LU decomposition or GMRES method. In particular, we can always discretize the initial curve $\Gamma^0=\vec{X}^0\in \mathcal{V}^{h,(r)}_{r_0(0),r_0(L_0)} \times \mathcal{V}^{h,(z)}_{r_0(0),r_0(L_0)}$ such that the mesh points $\{\vec X^0(\rho_j)\}_{j=0}^N$ are equally distributed along the curve with respect to the arc length, which means $|\partial_\rho\vec X^0|$ is a fixed constant. However, these mesh points at later time step will no longer be equally distributed. This is because the space domain we actually uniformly divide is the parameter domain $I=[0,1]$. The arc length of the curve is changing during time evolution, and there is no guarantee that the mesh points are moving accordingly such that the mesh quality is preserved. Therefore there is a high possibility that the quality of the mesh deteriorates, thus influencing the stability of the discrete scheme. Luckily, during our practical computation, we actually observe that the mesh quality is preserved quite well during the time evolution. This is because that Eq.~\eqref{eqn:fd3} can actually lead to a long-time mesh equi-distribution property, which has been fully investigated by J.W.~Barrett in~\cite{Barrett07ISO}.

\section{Numerical results and discussions}
Based on the PFEM presented above, we will present some numerical examples for simulating the solid-state dewetting with axis-symmetric geometry. In particular, the surface energy density is restricted to the $k$-fold as
\begin{equation}
\gamma(\theta)=1+\beta\cos(k\theta),\qquad0\leq\beta<\frac{1}{k^2-1}.
\end{equation}
Here $\beta$ represents the degree of the anisotropy, $k$ is the order of the rotational symmetry. Moreover, the mobility that determines the relaxation rate of the dynamical
contact angle to the equilibrium contact angle, is chosen numerically as $\eta=100$ for all cases. The detailed discussion about its influence to solid-state dewetting evolution process has been presented in ~\cite{Wang15}.
\subsection{Small islands}
\begin{figure}[!htp]
\centering
\includegraphics[width=1.0\textwidth]{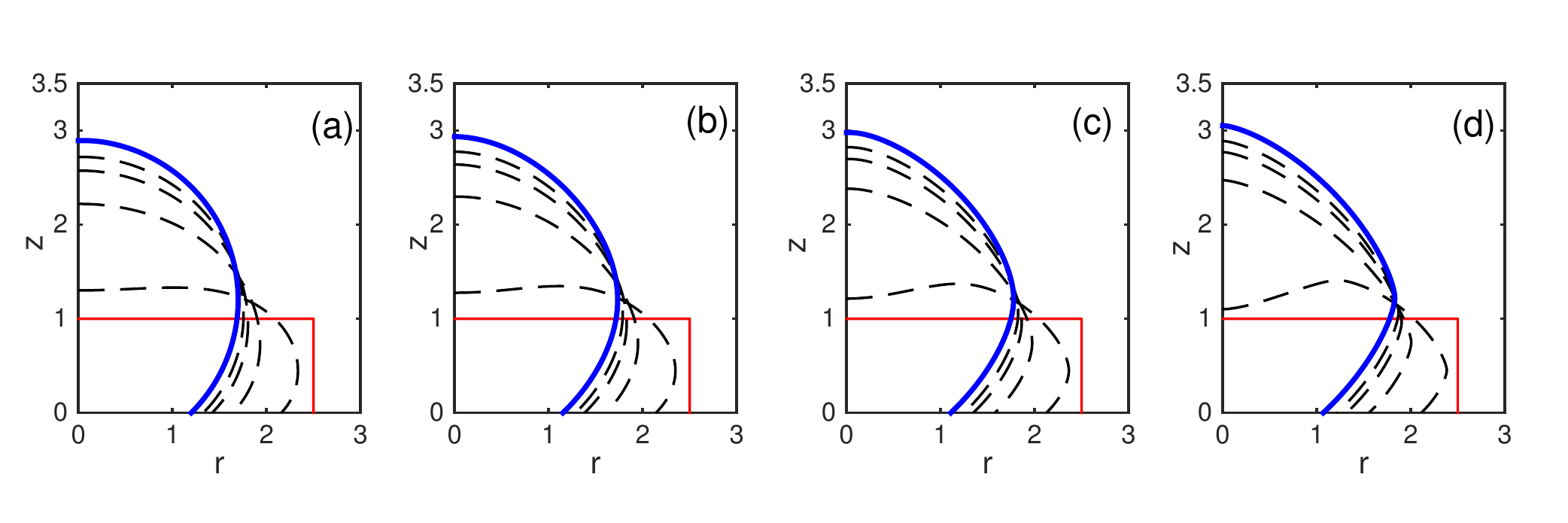}
\caption{Several snapshots of the curve in the axial direction during the evolution of an initially small cylinder towards its equilibrium for $4$-fold anisotropies with different $\beta$, $\sigma$ is fixed at $\cos(3\pi/4)$. (a) $\beta=0$; (b) $\beta=0.02$; (c) $\beta=0.04$; (d) $\beta=0.06$.}
\label{fig:det2}
\end{figure}
We first investigate the dynamic evolution of small cylindrical islands towards their equilibriums. The geometric evolutions of the islands are presented in Fig.~\ref{fig:det2} terms of the curves in the axial direction. The initial cylindrical islands (shown in red solid line) are chosen with height $z_0=1$ and radius $r_0=2.5$. The computational parameter $\sigma$ is fixed at $\cos(3\pi/4)$. For the surface anisotropy, the rotational symmetry $k$ is taken as $4$ but with $\beta$ increasing from $0$ to $0.06$ (Fig.~\ref{fig:det2}(a) - Fig.~\ref{fig:det2}(d)). The number of the mesh points during the numerical computation is chosen as $400$ and the time step is given as $\tau_m=0.0001$. From the figures, we can observe that in the four cases, the islands converge to their equilibriums (shown in blue solid line) and the contact angles converge to the equilibrium angles. As $\beta$ increases, the shapes of the equilibrium islands change from smooth to increasingly sharp corners. This indicates that the evolution of the curve in the axial direction for cylindrical islands is very similar to the evolution of rectangular islands in two dimensions (see~\cite{Wang15} for $2$D numerical examples). 

\begin{figure}[!htp]
\centering
\includegraphics[width=1\textwidth]{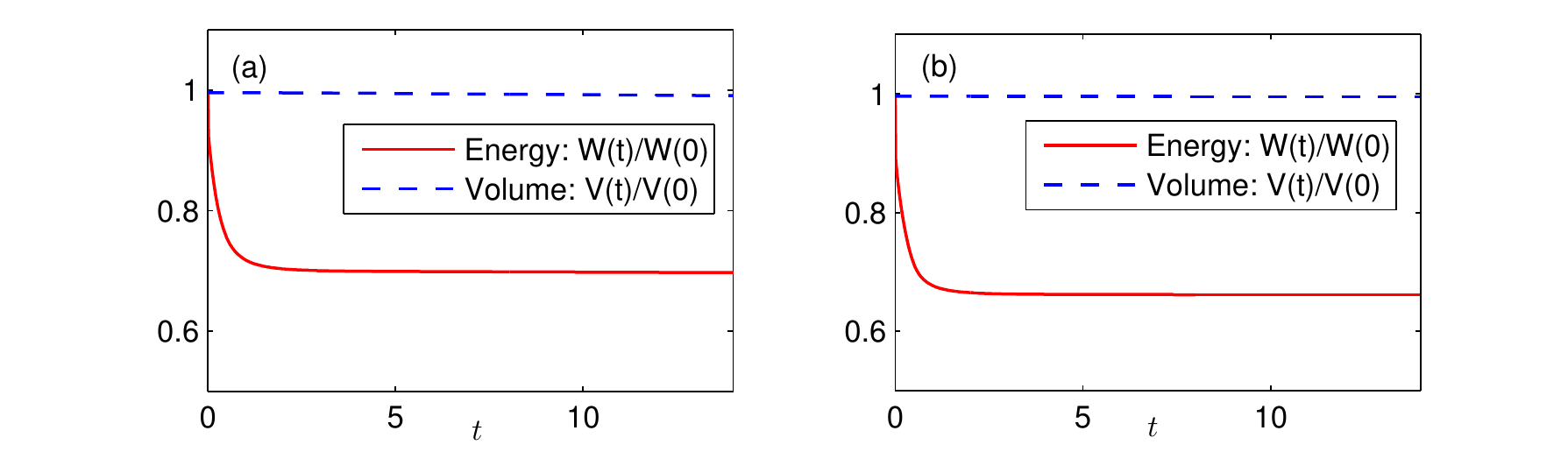}
\caption{The temporal evolution of the normalized total free energy and normalized volume with computational parameter chosen as $\sigma=\cos\frac{3\pi}{4}$. (a) Isotropic case with $\beta=0$; (b) anisotropic case with $\beta=0.06$.}
\label{fig:det4}
\end{figure}

We have proved theoretically that the total volume of the system is conserved and the total energy is dissipative during the time evolution for solid-state dewetting, see proposition.~\ref{prop:massconservation} and Proposition.~\ref{prop: energydissipation}.  Here we report that these properties can also be observed for the numerical discrete solutions. As can be seen in Fig.~\ref{fig:det4}, it depicts the temporal evolution of the normalized energy and volume for the thin film both for isotropic case (Fig.~\ref{fig:det4}(a)) and anisotropic case (Fig.~\ref{fig:det4}(d)). From the two figures, we can observe that the blue dash lines are almost horizontal for both cases, which means that the total volume is always conserved during the evolution. The trend of the red solid line indicates the dissipation for the total energy, and the total energy will stay unchanged as long as the islands obtain the equilibrium shape.

\begin{figure}[!htp]
\centering
\includegraphics[width=1\textwidth]{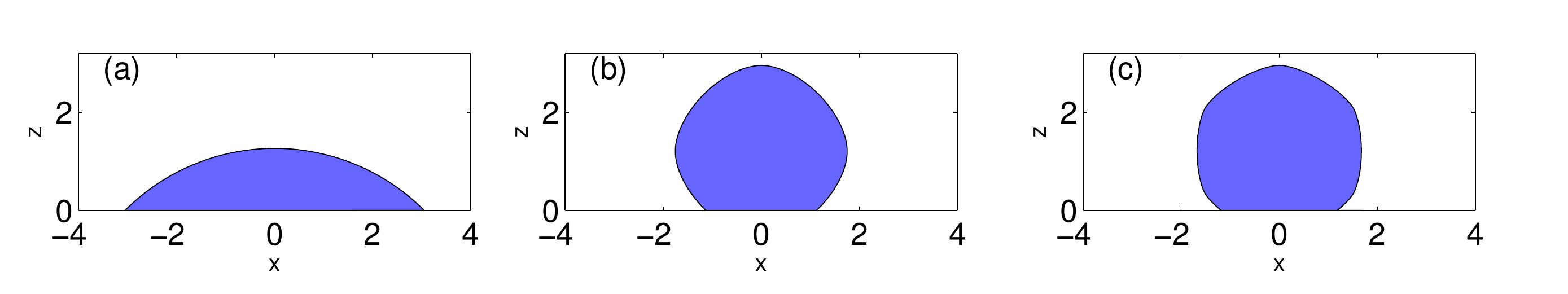}
\caption{The corresponding cross-section profile of the equilibrium shapes along $x$ axis with parameter chosen as (a) $\beta=0,\sigma=\cos\frac{\pi}{4}$; (b) $\beta=0.03,k=4,\sigma=\cos\frac{3\pi}{4}$;  (c) $\beta=0.02,k=6,\sigma=\cos\frac{3\pi}{4}$.}
\label{fig:det5}
\end{figure}

The surface energy density $\gamma(\theta)$ as well as the material constant $\sigma$ has great effects on the final equilibrium geometry. To observe the effects of the rotation symmetry $k$ and the $\sigma$, we perform some numerical simulations in Fig.~\ref{fig:det5}.  As can be seen, it shows the corresponding cross-section profile of the equilibrium shape along the $x$ direction. The initial thin film is taken as a cylindrical island with height $z_0=1$ and radius $r_0=2.5$. The parameter is chosen as following: for Fig.~\ref{fig:det5}(a), 
$\gamma(\theta)=1,\sigma=\cos\frac{\pi}{4}$; for Fig.~\ref{fig:det5}(b), $\gamma(\theta)=1+0.03\cos(4\theta),\sigma=\cos\frac{3\pi}{4}$; for Fig.~\ref{fig:det5}(c), $\gamma(\theta)=1+0.02\cos(6\theta),\sigma=\cos\frac{3\pi}{4}$. From these three numerical examples, we observe that $\sigma$ determines the contact angle while $k$ controls the number of sharp corners in the equilibrium shapes. 

\subsection{Large islands dynamics}
\begin{figure}[!htp]
\centering
\includegraphics[width=1\textwidth]{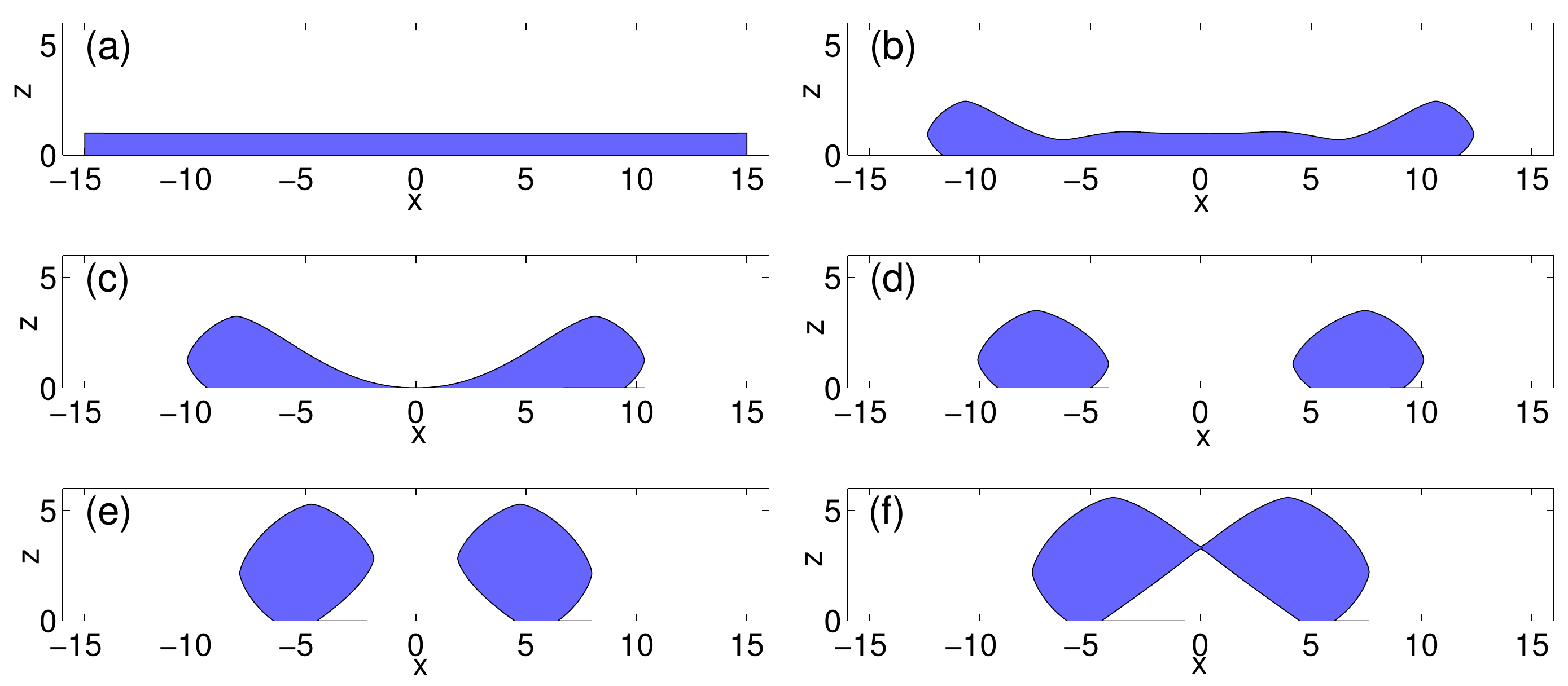}
\caption{The temporal geometry evolution of an initially cylindrical island with radius $r=15$ and height $z=1.0$, the parameter is chosen as $\sigma=\cos\frac{5\pi}{6},m=4,\beta=0.06$ where (a) $t=0$; (b) $t=10$; (c) $t=35.58$, the pinch off time; (d) $t=40$; (e)$t=220$; (f) $t=255.79$, the film forms a hole.}
\label{fig:det6}
\end{figure}
\begin{figure}[!htp]
\centering
\includegraphics[width=1.0\textwidth]{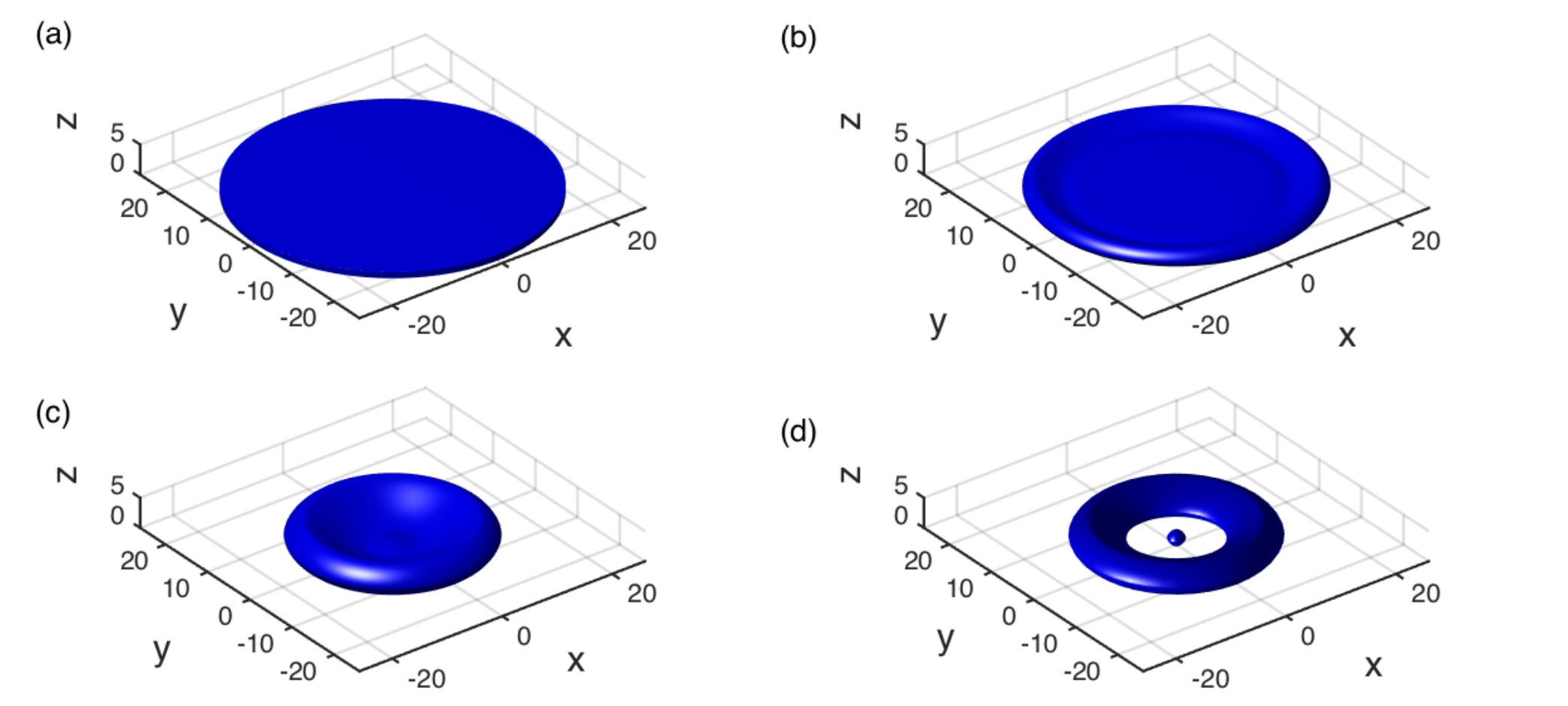}
\caption{The temporal geometry evolution of an initially cylindrical island with radius $r=25$ and height $z=1.0$, the parameter is chosen as $\sigma=\cos\frac{5\pi}{6},m=4,\beta=0.06$ where (a) $t=0$; (b) $t=10$; (c) $t=161.50$, the pinch off time; (d) $t=166.5$. }
\label{fig:det7a}
\end{figure}
Small cylindrical islands evolve and form single spherical (isotropic case) or near-spherical geometry with sharp corners (anisotropic case) as equilibrium shapes to obtain the minimizer of the total surface energy. For cylindrical islands with high ratio of radius to height, some new morphological features could be observed in the intermediate state of the islands towards the equilibrium. To investigate these features, we report the following numerical simulation example. The initial thin film is chosen as a large cylinder with height $z_0=1$ and radius $r_0=15$. The computational parameter are chosen as $\sigma=\cos\frac{5\pi}{6},k=4,\beta=0.06$. As illustrated in Fig.~\ref{fig:det6}, it cross-section profile along $x$-axis of the island geometry at some special moments. From Fig.~\ref{fig:det6}(a) to Fig.~\ref{fig:det6}(b), we can observe firstly that the rim and valley form near the outer contact lines of the thin film. As time evolves, the valley moves towards the centre, becomes deeper and finally touches the substrate. After the valley touches the substrate in Fig.~\ref{fig:det4}(c) at time $t=35.58$,  the new generated point quickly grows into an inner circular contact line and produces a thin film with toroidal geometry (shown in Fig.~\ref{fig:det4}(d)). This toroidal thin film then migrates towards the center and form a hole there (shown in Fig.~\ref{fig:det6}(f)). 

\renewcommand{\thefigure}{\arabic{figure} (Cont.)}
\addtocounter{figure}{-1}
\begin{figure}[!htp]
\centering
\includegraphics[width=1.0\textwidth]{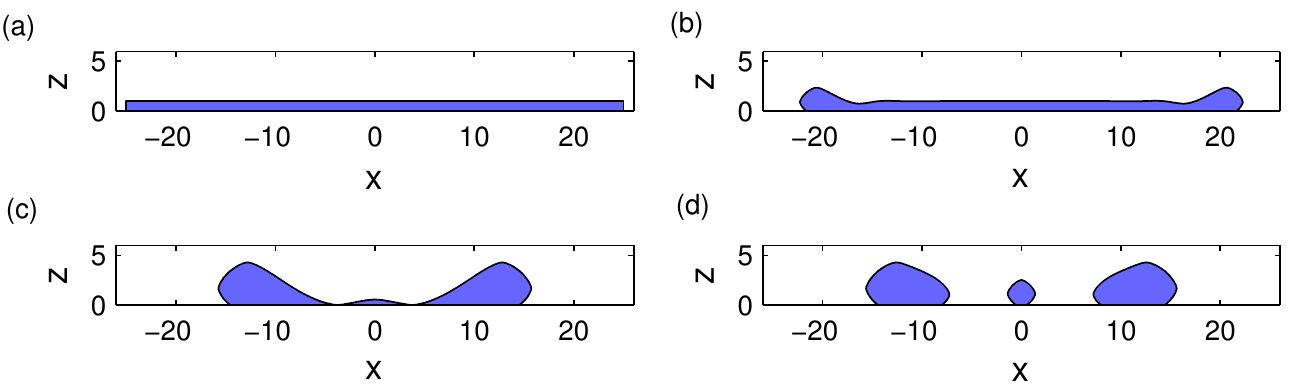}
\caption{The temporal evolution of an initially cylindrical island with radius $r=25$ and height $z=1.0$, the parameter is chosen as $\sigma=\cos\frac{5\pi}{6},m=4,\beta=0.06$ where (a) $t=0$; (b) $t=10$; (c) $t=161.50$, the pinch off time; (d) $t=166.5$;  This figure shows the corresponding cross-section profiles at each time.}
\label{fig:det7b}
\end{figure}
If we further increase the ratio of the radius to height for the initial thin film, we can observe that the cylindrical island will form a new contact line which separates the film into a single thin film in the center and a toroidal thin film outside. The three-dimensional morphological evolution of the large island and the corresponding cross-section profile along $x$-axis are shown in Fig.~\ref{fig:det7a} and Fig.~\ref{fig:det7b} respectively. The initial thin film is a cylinder with height $z_0=1$ and radius $r_0=25$. The computational parameters are chosen as $\gamma(\theta)=1+0.06\cos(4\theta)$ with $\sigma=\cos\frac{5\pi}{6}$. From the figures, we can observe that the large island eventually pinches off at time $t=161.5$ (Fig.~\ref{fig:det7a}(c)) and break up into a single small thin film in the centre and a toroidal thin film from the outside (Fig.~\ref{fig:det7a}(d)). Similarly, the toroidal thin film will shrink towards the center, but will not be shown here. 
\renewcommand{\thefigure}{\arabic{figure}}
\begin{figure}[!htp]
\centering
\includegraphics[width=0.45\textwidth]{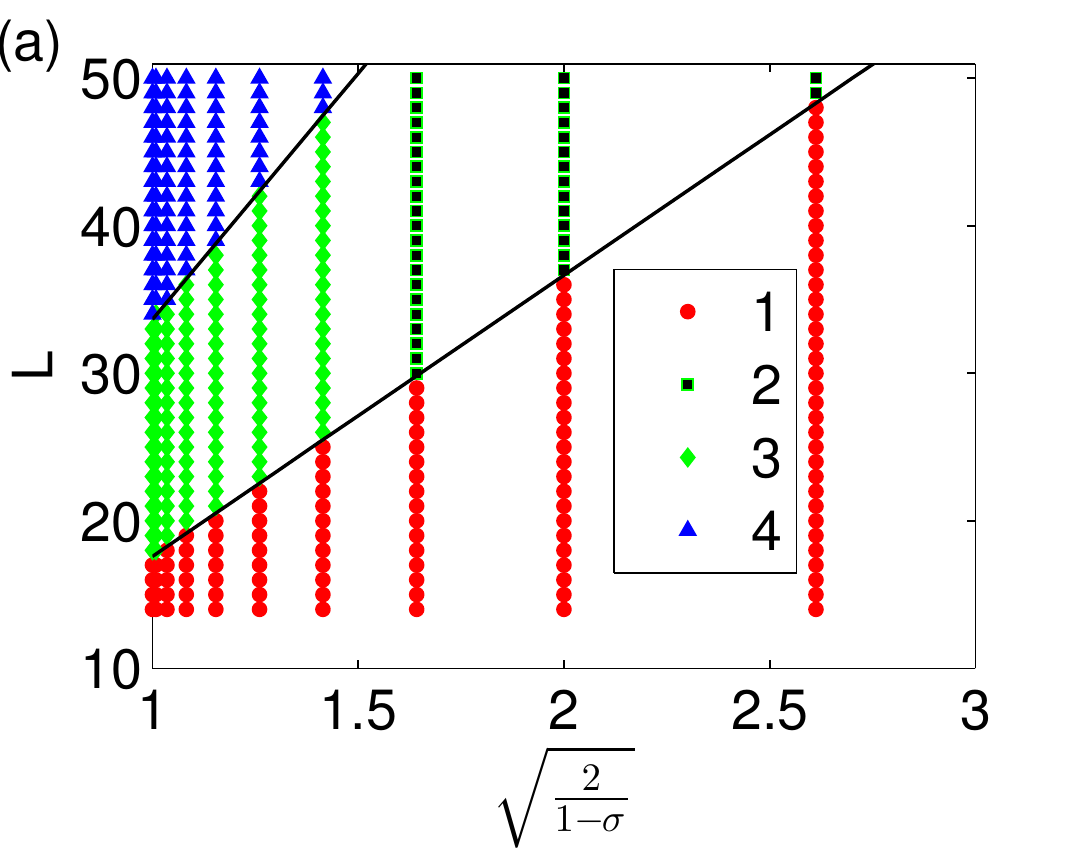}
\hspace{3ex}
\includegraphics[width=0.45\textwidth]{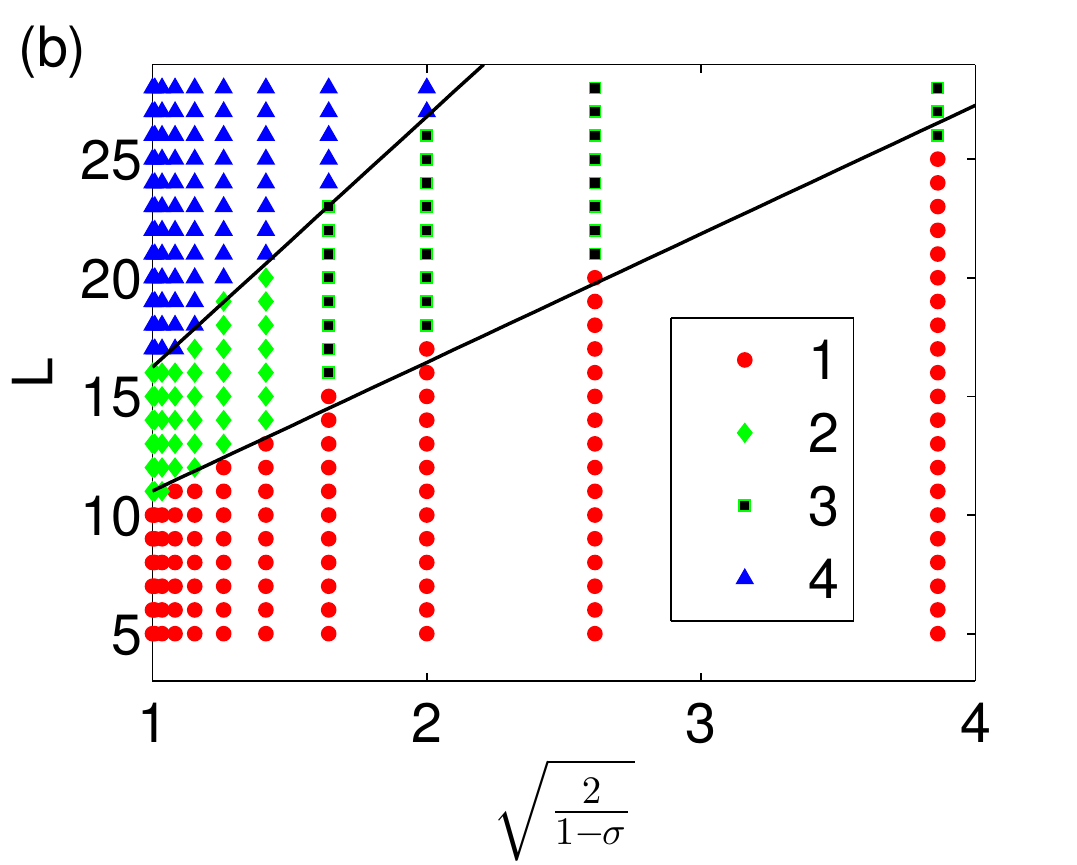}
\caption{Diagram showing the conditions ($L$ and $\sigma$, where the initial thin film is given as a cylinder with radius $L$ and height $1$,$\sigma$ is the material constant) for the occurrence of the four cases. (a) $\beta=0,m=4$, the equations for the two linear lines are: $L=19.06\sqrt{\frac{2}{1-\sigma}}-1.48$ and $L=33.35\sqrt{\frac{2}{1-\sigma}}+0.3055$ (b) $\beta=0.06,m=4$, the two equations for the two linear lines are $L=5.418\sqrt{\frac{2}{1-\sigma}}+5.60$ and $L=10.57\sqrt{\frac{2}{1-\sigma}}+5.66$. }
\label{fig:det8}
\end{figure}

Previous numerical examples have demonstrated three different types of evolution for an initial thin film with cylindrical geometry. The small cylindrical island can evolve into a single film as equilibrium shape, and during this evolution, no pinch-off events happen. However, larger islands could break up at the center to form a toroidal thin film (see Fig.~\ref{fig:det6}) or even leave a small single film at center (see Fig.~\ref{fig:det7a}) during this pinch-off. These different morphological evolutions are highly dependent on the initial ratio of the radius to height. Exactly, there must exist critical values that separate these three type evolutions. Moreover, when the large island breaks up into a toroidal thin film and then shrink towards the center, there are two possible evolutions at the time when the film merges itself, a void can be formed or no void is formed.  As illustrated in Fig.~\ref{fig:det8}, we have plotted the phase diagram both for isotropic and anisotropic surface energies, which shows clearly the effect of the parameter $\sigma$ and initial ratio of the cylinder on the evolution of the islands. In the figure, the four cases denoted by $1,2,3,4$ are used to represent different kinds of evolution. Here $1$: the island will form a single whole thin film for equilibrium shape; $2$: the island will pinch off at the center to form a toroidal thin film, and then the toroidal thin film will shrink toward $z$ axis and form void, see Fig.~\ref{fig:det4}; $3$: that the island will pinch off at the centre and then the toroidal thin film shrinks toward $z$ axis, but no void is formed for this case; $4$ The island will pinch off and break up into two parts, a single thin film at centre and a toroidal thin film outside, see Fig~\ref{fig:det5}. By observing these two figures, we find that both for isotropic and anisotropic surface energy, the phases can be separated by two linear curves, shown in Fig.~\ref{fig:det8}. This phenomenon has also be observed in two dimensions \cite{Wang15, Dornel06}. Besides, whether the void will form is highly dependent on the parameter $\sigma$. 

\section{Conclusions}
We firstly rigorously derived an axisymmetric sharp-interface model for solid-state dewetting based on the energy variational method. The proposed model describes the dynamic evolution of the film/vapor interface ($H^{-1}$ gradient flow) and migrations of the two contact lines along the flat substrate ($L^2$ gradient flow). Due to the cylindrical symmetry, the governing equations for the model are fourth-order geometric curve evolution PDEs about the axial curve with boundary conditions at the contact lines. Simultaneously, based on the first variation of the total surface energy, a mathematical description for the equilibrium shape is also presented. Secondly, we developed an efficient semi-implicit PFEM for solving the sharp-interface equations based on a good variational formulation. Thirdly, we investigated the complex morphological evolutions for solid-state dewetting by different numerical setups. In this paper, the first variation of the energy functional was performed based on a smooth axisymmetric perturbation of the interface, and this perturbation can actually collapse to a smooth vector-field perturbation of the axial open curve. Besides, the proposed PFEM shows some advantages over the previous approaches, e.g., the time step and mesh quality preservation. Moreover, from the numerical examples, we conclude that solid-state dewetting in three dimensions with cylindrical symmetry have exhibited a lot of interesting phenomena, such as the edge retraction, rim formations, pinch-off events, hole formations, shrinking instability and so on.  

The cylindrical symmetry we impose on the model actually is not always satisfied for the real solid-state dewetting problems. For example, the requirement for the symmetric geometry of the initial thin film seems difficult to be satisfied. Moreover, our approach failed to consider the perturbation effects along the azimuthal direction. Physical experiments by E.~Pairam \cite{Pairam09} show that toroidal droplets can break up into a precise number of small droplets and only the fat toroidal droplets are able to migrate towards the center and merge before they break up along the azimuthal direction, depending on the thickness of the torus relative to its circumference. Experiments for solid-state dewetting ~\cite{Ye11b} also demonstrates that square ring patches of single crystals nickels can break up into small particles along the ring due to the Rayleigh instability. This demonstrates that the Rayleigh instability in the azimuthal direction and shrinking instability in the axial direction are competing with each other to determine the dynamics geometries of the torus (com-
petition of the two-time scales: one for migration towards
center and one for pinch-off along azimuthal direction). Thus the symmetric model prevents us from understanding well the full kinetic patterns and morphological characteristics for the solid-state dewetting problem, and it inspires us to work on the full $3$D problems in the future. However, the axisymmetric model, as well as the proposed PFEM, can shed some light on the further research for the real three-dimensional problem.   

\section*{Acknowledgements}
The author would like to thank Prof.~Weizhu Bao, Prof.~Wei Jiang for fruitful discussions. We acknowledge the support from the Ministry of Education of
Singapore grant R-146-000-247-114. 
\section*{References}
\bibliographystyle{elsarticle-num}
\bibliography{thebib}

\end{document}